\documentclass[format=sigconf,natbib=false,anonymous=false,screen]{acmart}

\def\myauthors{David Otero and Javier Parapar}
\def\mytitle{Hybrid Pooling with LLMs via Relevance Context Learning}

\AtBeginDocument{%
  }

\copyrightyear{2026}
\acmYear{2026}
\setcopyright{cc}
\setcctype{by}
\acmConference[SIGIR '26]{Proceedings of the 49th International ACM SIGIR Conference on Research and Development in Information Retrieval}{July 20--24, 2026}{Melbourne, VIC, Australia}
\acmBooktitle{Proceedings of the 49th International ACM SIGIR Conference on Research and Development in Information Retrieval (SIGIR '26), July 20--24, 2026, Melbourne, VIC, Australia}
\acmDOI{10.1145/3805712.3809669}
\acmISBN{979-8-4007-2599-9/2026/07}

\usepackage{multirow}
\usepackage{soul}
\usepackage{subcaption}
\usepackage{amsmath}
\usepackage{graphicx}
\usepackage{url}
\usepackage[inline]{enumitem}
\usepackage{acronym}
\usepackage{booktabs}
\usepackage{longtable}
\usepackage{pdflscape}
\usepackage{tabularx}
\usepackage{afterpage}
\usepackage{siunitx}
\usepackage{algpseudocode, algorithm}
\usepackage{calc}
\usepackage[backend=biber,style=acmnumeric,datamodel=acmdatamodel,backref=true,backreffloats=false,sortcites]{biblatex}
\usepackage{datenumber}
\usepackage{colortbl}
\usepackage{cleveref}
\usepackage{pifont}
\usepackage{fontawesome5}
\hypersetup{
  pdfauthor={\myauthors},
  pdftitle={\mytitle},
  pdfsubject={Information Retrieval},
}

\graphicspath{{plots/}}

\DefineBibliographyStrings{american}{%
  backrefpage={cited on p.},
  backrefpages={cited on pp.}
}

\algblock{Input}{EndInput} 
\algnotext{EndInput}  
\algblock{Output}{EndOutput} 
\algnotext{EndOutput} 

\definecolor{observationbg}{RGB}{255,248,220}  
\definecolor{observationtext}{RGB}{139,69,19}  

\newcommand{\un}[1]{\underline{#1}}

\author{David Otero}
\email{david.otero.freijeiro@udc.es}
\orcid{0000-0003-1139-0449}
\affiliation{
  \institution{IRLab, CITIC, Universidade da Coruña}
  \city{A Coruña}
  \streetaddress{Facultad de Inform\'{a}tica, Campus de Elvi\~{n}a s/n}
  \country{Spain}
  \postcode{15071}
}

\author{Javier Parapar}
\email{javier.parapar@udc.es}
\orcid{0000-0002-5997-8252}
\affiliation{
  \institution{IRLab, CITIC, Universidade da Coruña}
  \city{A Coruña}
  \streetaddress{Facultad de Inform\'{a}tica, Campus de Elvi\~{n}a s/n}
  \country{Spain}
  \postcode{15071}
}

\begin{document}

\title[\mytitle]{\mytitle}

\begin{abstract}

High-quality relevance judgements over large query sets are essential for evaluating Information Retrieval (IR) systems, yet manual annotation remains costly and time-consuming. Large Language Models (LLMs) have recently shown promise as automatic relevance assessors, but their reliability is still limited. Most existing approaches rely on zero-shot prompting or in-context learning (ICL) with a small number of labelled examples. However, standard ICL treats examples as independent instances and fails to explicitly capture the underlying relevance criteria of a topic, restricting its ability to generalise to unseen query-document pairs. To address this limitation, we introduce Relevance Context Learning (RCL), a novel framework that leverages human relevance judgements to explicitly model topic-specific relevance criteria. Rather than directly using labelled examples for in-context prediction, RCL first prompts an LLM (Instructor LLM) to analyse sets of judged query-document pairs and generate explicit narratives that describe what constitutes relevance for a given topic. These relevance narratives are then used as structured prompts to guide a second LLM (Assessor LLM) in producing relevance judgements. To evaluate RCL in a realistic data collection setting, we propose a hybrid pooling strategy in which a shallow depth-\textit{k} pool from participating systems is judged by human assessors, while the remaining documents are labelled by LLMs. Experimental results demonstrate that RCL substantially outperforms zero-shot prompting and consistently improves over standard ICL. Overall, our findings indicate that transforming relevance examples into explicit, context-aware relevance narratives is a more effective way of exploiting human judgements for LLM-based IR dataset construction.

\end{abstract}


\maketitle


\section{Introduction}
\label{sec:intro}

\sloppy Cranfield-style test collections~\cite{Cleverdon1967} remain the cornerstone of rigorous offline evaluation for Information Retrieval (IR) systems~\cite{Sanderson2010}. Such collections comprise a document corpus, a set of queries (or topics), and relevance assessments that associate each query-document pair with a human-assigned relevance label (\textit{qrels}). While this paradigm has enabled decades of systematic progress in IR evaluation, the construction of high-quality relevance assessments remains exceptionally costly and time-consuming. Given that modern test collections often contain millions of documents, exhaustively judging all query-document pairs is infeasible. As a result, pooling strategies, most commonly depth-\textit{k} pooling, are employed, whereby only a subset of documents retrieved by a diverse set of systems is judged~\cite{Voorhees2001a,Buckley2005}. Nevertheless, despite these costs and practical challenges~\cite{Buckley2007}, manual relevance assessment of pooled documents remains the primary bottleneck in the construction of large-scale test collections.

The high cost of human annotation has motivated extensive research on cost-reduction strategies for relevance assessment and collection building. Prior work has explored active learning approaches to identify the most informative documents for judgement~\cite{Cormack2018}, multi-armed bandit methods to allocate labelling effort across queries or systems~\cite{Losada2016a,Lipani2016,Losada2017a,Lipani2019}, and improved pooling and sampling strategies that aim to preserve evaluation quality while reducing annotation effort~\cite{Otero2023a}. In parallel, crowdsourcing has been investigated as a means of trading off cost and judgement quality~\cite{Alonso2012}. While these methods reduce reliance on expert assessors, they still fundamentally depend on human judgements and face scalability challenges for very large, dynamic, or continuously evolving corpora.

\sloppy Recently, advances in Large Language Models (LLMs) have opened up the possibility of partially or fully automating the generation of relevance assessments, gaining considerable attention within the IR community~\cite{Alaofi2026,Balog2025,Otero2025b,Merlo2025,McKechnie2025a,Alaofi2024,Upadhyay2024c,Faggioli2023,Upadhyay2024b,Takehi2024,Soboroff2024,Upadhyay2024a,Abbasiantaeb2024,Thomas2024,MacAvaney2023}. Such works explore the use of LLMs to reduce the dependency on costly human assessors. Several studies have demonstrated that LLMs can achieve high correlation with human judgements in relevance assessment tasks~\cite{Upadhyay2024c}, although not without limitations~\cite{Otero2025b,Soboroff2024}. Typically, these approaches deploy zero-shot prompting strategies, where only the task is described, or in-context learning (ICL), where the model is provided with a few examples to demonstrate the assessment task~\cite{McKechnie2025a}.

However, existing LLM-based approaches face several challenges. First, the performance of ICL-based methods can be sensitive to the selection and position of examples~\cite{Liu2024}, and since LLMs have limited attention spans, indefinitely increasing the number of examples does not necessarily improve performance. More fundamentally, standard ICL treats each example as an isolated query-document pair and does not explicitly capture the notion of relevance underlying a given topic. This generalisation problem~\cite{Dong2024} reduces the ability of the model to work well beyond the specific examples provided, as it lacks an explicit understanding of what makes a document relevant for a particular information need.

\begin{figure}
    \centering
    \includegraphics[width=\linewidth]{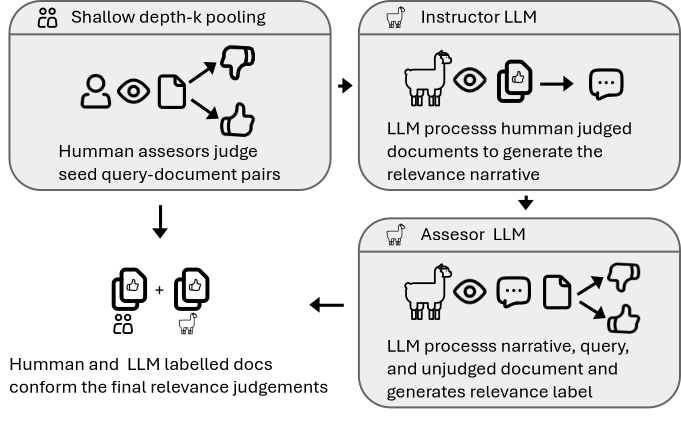}
    \Description{The hybrid pooling approach using Relevance Context Learning framework. The process begins with a \textit{shallow depth-$k$ pool} where human assessors provide initial relevance judgements. These judgements are fed into the \textit{Instructor LLM}, which analyses the query--document pairs to synthesise explicit \textit{Relevance Narratives}. These narratives serve as structured prompts for the \textit{Assessor LLM}, which labels the remaining documents in the pool. The final \textit{Hybrid Pooled Data} combines human and LLM-generated labels to create a cost-effective, high-quality IR evaluation dataset.}
    \caption{The hybrid pooling approach using Relevance Context Learning framework. The process begins with a \textit{shallow depth-$k$ pool} where human assessors provide initial relevance judgements. These judgements are fed into the \textit{Instructor LLM}, which analyses the query--document pairs to synthesise explicit \textit{Relevance Narratives}. These narratives serve as structured prompts for the \textit{Assessor LLM}, which labels the remaining documents in the pool. The final \textit{Hybrid Pooled Data} combines human and LLM-generated labels to create a cost-effective, high-quality IR evaluation dataset.}
    \label{fig:rcl-flow}
\end{figure}

To address these limitations, we propose Relevance Context Learning (RCL) (see \Cref{fig:rcl-flow}), a novel approach that leverages a small set of human-annotated examples to learn the relevance context of a topic. Rather than directly using labelled examples when judging query-document pairs, RCL operates in two stages. First, we prompt an LLM (Instructor LLM) to analyse the human-annotated examples and generate an explicit explanation of what constitutes relevance for the given topic. This explanation captures the underlying relevance criteria as natural-language narratives and assessment instructions. Second, we use this generated relevance explanation in a prompt to guide another LLM (Assessor LLM) when judging new query-document pairs. By explicitly modelling the relevance context, RCL enables the LLM to better generalise to unseen documents while maintaining interpretability. To evaluate this approach in a realistic setting, we propose a \textit{hybrid pooling strategy} in which a very shallow depth-\textit{k} pool from participating systems is judged by human assessors, while the remainder of the pool is labelled by LLMs.

Our experiments\footnote{Code available at \faGithub~\url{https://github.com/IRLab-UDC/rcl}} on three standard test collections (TREC Deep Learning 2019 and 2020, and TREC-8) demonstrate that RCL achieves competitive or superior effectiveness compared to standard ICL. Beyond effectiveness, RCL offers clear practical advantages in terms of cost and inference efficiency. In RCL, relevance judgements are conditioned on a compact, topic-specific relevance narrative, which requires substantially fewer input tokens than ICL approaches that rely on multiple full query-document examples. In contrast, example-based ICL often involves lengthy document passages and repeated query context, resulting in large prompts that not only strain the model’s attention capacity but also increase inference latency and computational cost. These advantages are particularly pronounced on collections with long documents, such as TREC-8, where including multiple full-document examples quickly exhausts the LLM’s context window. In such settings, narrative-based prompting substantially outperforms example-based ICL, highlighting the benefits of abstracting relevance criteria into a concise and reusable form. We further investigate hybrid configurations that combine relevance narratives with in-context examples. Overall, our findings support hybrid pooling with RCL as a cost-effective, scalable, and operationally efficient approach to constructing high-quality IR judgements.


\section{Hybrid Pooling for LLM-based Judgement}
\label{sec:hybrid-pooling}

Prior work on LLM-based relevance assessment typically starts from existing official qrels files and evaluates how well LLMs can replicate human judgements \cite{Faggioli2023,Balog2025,McKechnie2025a}. While these studies provide valuable insights into LLM capabilities, they do not address the practical scenario of using LLMs to construct a new test collection from scratch. To the best of our knowledge, prior work has not provided a complete, directly deployable pipeline for real-world collection building that starts from participating systems in a TREC-style campaign rather than from complete qrels.

Our first contribution addresses this gap by proposing a \emph{hybrid pooling strategy} that mirrors the realistic workflow of test collection construction. Starting only from the ranked lists submitted by participating systems, we divide the annotation workload between human assessors and LLMs in a way that maximizes the utility of limited human effort while leveraging LLMs to extend coverage.

\subsection{Method}
\label{sec:hybrid-pooling-method}

Our approach follows the standard pooling methodology used in evaluation campaigns such as TREC, but with a crucial modification. Instead of having human assessors judge the entire pool of documents, we divide the annotation workload between humans and LLMs. Specifically, we employ a depth-$k$ pooling strategy where:

\begin{enumerate}
    \item We collect the top-$k$ documents from each participating retrieval system to form a pool of candidate documents for each topic.
    \item Human assessors judge only a shallow subset of this pool, specifically the top-$k_{human}$ documents (where $k_{human} < k$).
    \item The remaining documents in the pool (from rank $k_{human}+1$ to $k$) are judged by an LLM.
\end{enumerate}

This hybrid strategy reflects a practical scenario where annotation budget is limited and we want to maximize the impact of human effort by focusing it on the most likely relevant documents (those ranked highest by multiple systems), while using LLMs to extend coverage to deeper ranks. The human-judged subset serves two purposes: it provides high-quality judgements for the most important documents, and it supplies the training signal that allows the LLM to learn the relevance criteria for each topic. 

The complete process for constructing hybrid qrels is formalised in Algorithm~\ref{alg:qrels}. Starting from the ranked lists submitted by participating systems, we create a pool for each topic and divide it into shallow and deep portions based on the pooling depth $k_{human}$. Human assessors judge the shallow portion, while the deeper portion is passed to the LLM for automatic assessment. The LLM judgement process (Algorithm~\ref{alg:llmjudge}) operates by selecting relevant examples from the human-judged set for each document to be assessed and using these examples to guide the LLM through in-context learning. We investigate several strategies for example selection, including sampling from both relevant and non-relevant judgements or restricting selection to a single relevance class. Further details on these strategies are provided in Section~\ref{sec:hybrid-pooling-setup}.

\begin{algorithm}
\caption{Constructing hybrid qrels from participating systems}
\label{alg:qrels}
\begin{algorithmic}[1]
\Require System runs $\mathcal{S} = \{S_1, S_2, \ldots, S_n\}$, topics $\mathcal{Q}$, pooling depths $k$ and $k_{human}$ (where $k_{human} < k$)
\Ensure Final qrels $R_{final}$

\State Initialize $R_{human} \leftarrow \emptyset$, $P_{llm} \leftarrow \emptyset$

\For{each topic $q \in \mathcal{Q}$}
    \State $Pool_q \leftarrow$ CreatePool($\mathcal{S}, q, k$) \Comment{Collect top-$k$ docs from each system}
    \State $Pool_{shallow} \leftarrow$ documents in $Pool_q$ at ranks $\leq k_{human}$
    \State $Pool_{deep} \leftarrow$ documents in $Pool_q$ at ranks $> k_{human}$

    \State $R_{human} \leftarrow R_{human} \cup$ HumanJudge($q, Pool_{shallow}$)
    \State $P_{llm} \leftarrow P_{llm} \cup \{(q, d) : d \in Pool_{deep}\}$
\EndFor

\State $R_{llm} \leftarrow$ LLMJudge($R_{human}, P_{llm}$) \Comment{Algorithm~\ref{alg:llmjudge}}
\State $R_{final} \leftarrow R_{human} \cup R_{llm}$

\State \Return $R_{final}$

\end{algorithmic}
\end{algorithm}

\begin{algorithm}
\caption{LLMJudge: In-Context Learning for IR Relevance Judgements}
\label{alg:llmjudge}
\begin{algorithmic}[1]
\Require Human-judged set $R_{human}$, Candidate pairs $P_{llm}$, LLM $\Phi$, number of examples $k$, selection strategy $S$
\Ensure LLM-generated relevance labels $R_{llm}$

\State Initialize $R_{llm} \leftarrow \emptyset$

\For{each $(q,d) \in P_{llm}$}
    \State $C \leftarrow$ SelectExamples$(R_{human}, q, d, k, S)$ \Comment{Select $k$ examples using strategy $S$}
    \State Construct prompt $P_{prompt}(q,d,C)$ containing task instructions, $C$, and $(q,d)$
    \State $\hat{r} \leftarrow \Phi(P_{prompt})$ \Comment{LLM generates judgement}
    \State Add $(q,d,\hat{r})$ to $R_{llm}$
\EndFor

\State \Return $R_{llm}$

\end{algorithmic}
\end{algorithm}

To evaluate this method, we compare it against common ICL approaches as well as against the stratified sampling strategy proposed by McKechnie et al.~\cite{McKechnie2025a}, which, to the best of our knowledge, is the only prior work that has evaluated example selection strategies for ICL-based relevance assessment.

\subsection{Experimental Setup}
\label{sec:hybrid-pooling-setup}

\subsubsection{Datasets}

We evaluate our hybrid pooling approach on three standard IR test collections: TREC Deep Learning passage collections DL-19~\cite{Craswell2019} and DL-20~\cite{Craswell2020}, and TREC-8~\cite{Voorhees2000}.

\subsubsection{Pooling Strategies}

We compare two strategies for selecting which documents human assessors should judge:

\begin{itemize}
    \item \textbf{Hybrid shallow depth-$k$ pooling}: Human assessors judge the top-3 documents from each participating system. This simulates the realistic scenario where annotation effort is focused on the documents most likely to be relevant. For all the experiments performed in our work, we set $k = 10$ and $k_{human} = 3$
    \item \textbf{Stratified sampling}~\cite{McKechnie2025a}: Human assessors judge a 10\% stratified sample of the final pool, \textbf{with respect to relevance distribution} in the complete qrels.
\end{itemize}

Both strategies result in comparable human annotation effort, allowing for a fair comparison.

\subsubsection{Model Configuration}

We use \texttt{Llama-3.1-8B-Instruct} as the LLM for generating relevance judgements for a fair comparison against past work~\cite{McKechnie2025a}. Model inference is performed with vLLM\footnote{\url{https://docs.vllm.ai/en/latest}}, a high-performance inference framework that enables efficient processing of large batches. Structured outputs are enforced using grammar-constrained generation, ensuring the model can only respond with ``Relevant'' or ``Not Relevant''. Each model response is mapped to a binary relevance label (Relevant $\rightarrow$ 1, Not Relevant $\rightarrow$ 0).

\subsubsection{Example Selection Strategies}

We compare several approaches for selecting which examples from the human-judged pool should be provided to the LLM as in-context demonstrations:

\begin{itemize}
    \item \textbf{Zero-shot}: The LLM receives only task instructions without any examples.
    \item \textbf{ICL}: Examples are randomly selected from the human-judged pool, regardless of their relevance label.
    \item \textbf{ICL Relevant}: Examples are randomly selected from only the relevant documents in the human-judged pool.
\end{itemize}

We vary the number of examples ($k \in \{1, 2, 3\}$) to understand how performance scales with the amount of context provided.

\subsubsection{Evaluation Metrics} We use AP@1000\footnote{We also have results using nDCG@10, but limit our discussion to AP due to space limitations. Trends are the same.} for scoring the runs and we evaluate the quality of LLM-generated judgements using:
\begin{itemize}
    \item \textbf{Averaged per-query F1} against the gold relevance labels, which measures the accuracy of binary relevance predictions.
    \item \textbf{Matthews Correlation Coefficient (MCC)}: a balanced classification metric that we use for evaluating the preservation of correct decisions regarding the significance differences between pairs of runs~\cite{McKechnie2025b}. We use a two-sided Wilcoxon test with Benjamini-Hochberg for correction, as recommended in previous work~\cite{Otero2025a}.
\end{itemize}

\subsection{Results}
\label{sec:hybrid-pooling-results}

\begin{table*}
\centering
\setlength{\tabcolsep}{3pt}
\footnotesize

\caption{Results for F1 and MCC on DL-19, DL-20, and TREC-8. Results on the DL collections are averaged over 10 runs. Best values for each collection and metric are shown in \textsc{bold}, while the second best are \un{underlined}. F1 scores for the human-only pooling are not reported, as only a small subset was examined and those labels are used as gold judgements, yielding a trivial F1 of 1 on the evaluated subset. Human assessors examined the top-3 documents from participating systems in the Depth-\textit{k} approach, and a 10\% stratified sample (with respect to gold relevance in the final \textit{qrels}) in the Sampling approach, resulting in comparable human assessment effort in both cases.}
\label{tab:exp1}

\begin{tabular}{l|c|cc|cc|cc|cc|cc|cc}

\toprule

\multicolumn{2}{l}{} & \multicolumn{4}{|c|}{\textbf{DL-19}} & \multicolumn{4}{c|}{\textbf{DL-20}} & \multicolumn{4}{c}{\textbf{TREC-8}} \\

\midrule

\multicolumn{2}{l}{} & \multicolumn{2}{|c|}{\textbf{F1}} & \multicolumn{2}{c|}{\textbf{MCC}} & \multicolumn{2}{c|}{\textbf{F1}} & \multicolumn{2}{c|}{\textbf{MCC}} & \multicolumn{2}{c|}{\textbf{F1}} & \multicolumn{2}{c}{\textbf{MCC}} \\

\midrule

\multicolumn{1}{l}{\textbf{Strategy}} & \textbf{Shots} & \multicolumn{1}{c}{\textbf{Sampl.}} & \multicolumn{1}{c|}{\textbf{Depth-\textit{k}}} & \multicolumn{1}{c}{\textbf{Sampl.}} & \multicolumn{1}{c|}{\textbf{Depth-\textit{k}}} & \multicolumn{1}{c}{\textbf{Sampl.}} & \multicolumn{1}{c|}{\textbf{Depth-\textit{k}}} & \multicolumn{1}{c}{\textbf{Sampl.}} & \multicolumn{1}{c|}{\textbf{Depth-\textit{k}}} & \multicolumn{1}{c}{\textbf{Sampl.}} & \multicolumn{1}{c|}{\textbf{Depth-\textit{k}}} & \multicolumn{1}{c}{\textbf{Sampl.}} & \multicolumn{1}{c}{\textbf{Depth-\textit{k}}} \\

\midrule

\arrayrulecolor{black!30}

\textsc{Baseline} (Human assessor judgements) & - & - & - & 0.4891 & 0.7316 & - & - & 0.4699 & 0.7635 & - & - & 0.5866 & \textbf{0.8284} \\

\cmidrule{1-14}

\textsc{Zero Shot} & 0 & 0.5726 & 0.7598 & 0.2438 & 0.6092 & 0.5426 & 0.6628 & 0.15555 & 0.6754 & 0.3061 & 0.5154 & 0.4137 & 0.5096 \\

\cmidrule{1-14}

\multirow{3}{*}{\textsc{ICL}}
& 1 & 0.6956 & 0.8353 & 0.6656 & 0.8847 & 0.6675 & 0.7784 & 0.4580 & 0.8181 & 0.5347 & 0.6967 & 0.4952 & 0.6244 \\
& 2 & 0.7033 & 0.8541 & 0.6769 & 0.9134 & 0.6874 & 0.8070 & 0.4592 & 0.8373 & 0.5589 & 0.7097 & 0.5045 & 0.6460 \\
& 3 & 0.7095 & 0.8559 & 0.6633 & 0.8890 & 0.6909 & 0.8160 & 0.4647 & 0.8365 & 0.5772 & 0.7272 & 0.5193 & 0.6606 \\

\cmidrule{1-14}

\multirow{3}{*}{\textsc{ICL Relevant}}
& 1 & 0.7483 & 0.8654 & 0.7767 & 0.9144 & 0.7294 & 0.8384 & 0.6726 & 0.8656 & 0.6157 & \textbf{0.7677} & 0.5271 & \un{0.7269} \\
& 2 & 0.7387 & \un{0.8685} & 0.7789 & \textbf{0.9270} & 0.7309 & \un{0.8405} & 0.6293 & \textbf{0.8785} & 0.5860 & \un{0.7409} & 0.5739 & 0.7055 \\
& 3 & 0.7381 & \textbf{0.8753} & 0.7719 & \un{0.9236} & 0.7383 & \textbf{0.8489} & 0.6066 & \un{0.8784} & 0.5780 & 0.7315 & 0.5720 & 0.7047 \\

\arrayrulecolor{black}

\bottomrule

\end{tabular}
\end{table*}

\Cref{tab:exp1} compares the performance of ICL strategies under both stratified sampling and depth-$k$ pooling approaches. The results reveal several important insights about practical test collection construction.

\subsubsection{Depth-$k$ vs. Stratified Sampling}
Depth-$k$ hybrid pooling consistently outperforms stratified sampling across all collections. Using ICL Relevant (3 shots), depth-$k$ achieves an F1 of 0.8753 on DL-19 compared to 0.7381 with stratified sampling, a relative improvement of about 19\%. Similar gains are observed on DL-20 (0.8489 vs.\ 0.7383, about 15\%) and TREC-8 (0.7315 vs.\ 0.5780, about 27\%). These improvements are also reflected in MCC, where depth-$k$ yields substantially higher discriminative power (e.g., 0.9236 vs.\ 0.7719 on DL-19, about 20\% relative improvement).

These results align with recent evidence showing that LLMs can approximate human relevance assessment at an aggregate, system-ranking level. For example, Upadhyay et al.\ report that UMBRELA~\cite{Upadhyay2024c} achieves a Kendall’s $\tau$ correlation of approximately 0.89 with NIST system rankings on the TREC RAG 2024 dataset---often considered close to the ceiling of inter-assessor agreement among humans. While such findings do not imply that LLMs can replace human assessors at the level of individual judgements~\cite{Clarke2024}, they provide strong empirical support for using LLMs in pooling and system-level evaluation tasks.

At the same time, Clarke et al.~\cite{Clarke2024} highlight the risks of relying on \emph{pure} LLM-based relevance assessment. In the TREC 2024 RAG Track, the WaterlooClarke submission (run \texttt{uwc1}) deliberately constructed a ranking by re-ranking pooled documents using an LLM prompted to act as a relevance judge. Under UMBRELA evaluation, this run ranked 5th overall, whereas under human evaluation it dropped sharply to 28th. This discrepancy indicates a form of circularity: when both the ranker and the judge rely on similar LLM capabilities, they may reinforce superficial relevance signals that deviate from human notions of utility and correctness. As a result, high average correlation does not guarantee robustness in adversarial or self-referential settings.

In this context, the superiority of depth-$k$ hybrid pooling further reinforces the need for human oversight at critical decision points. By anchoring evaluation with human judgements at shallow ranks---where systems compete most intensely---hybrid depth-$k$ pooling mitigates circularity effects while still leveraging LLMs to scale assessment to deeper ranks. In contrast, stratified sampling distributes limited human effort more uniformly across the pool, including many low-impact documents, which weakens supervision where it matters most for ranking-sensitive evaluation.

\subsubsection{Limitations of Standard ICL}

Standard ICL is sensitive to both example selection and prompt length. Using only relevant examples consistently outperforms random mixtures of relevant and non-relevant examples. While performance improves when increasing from one to two examples, adding more examples often leads to diminishing or even negative returns, particularly for collections with long documents (e.g., TREC-8), where the LLM’s context window is quickly exhausted. These effects are consistent with recent findings on \emph{lost-in-the-middle} and order sensitivity in long-context reasoning, which show that LLMs struggle to attend uniformly across long inputs and may underutilize examples placed away from the beginning or end of the prompt~\cite{Liu2024,Liao2025}.

The comparison between ICL and ICL Relevant is particularly instructive. On DL-19 with three shots under depth-$k$ pooling, basic ICL achieves 0.8559 F1, whereas ICL Relevant reaches 0.8753 F1, corresponding to a 2.3\% improvement. The gap is even larger in terms of MCC (0.8890 vs.\ 0.9236, a 3.9\% improvement), indicating that restricting examples to relevant documents produces judgements with stronger discriminative power. These results suggest that both the content and ordering of examples play a critical role in ICL effectiveness, and that increasing prompt length alone does not guarantee improved performance.

Zero-shot performance provides an important baseline, illustrating what LLMs can achieve without any topic-specific guidance. Across all collections, zero-shot prompting performs substantially worse than ICL-based approaches. The gap is most pronounced on TREC-8, where zero-shot achieves 0.5154 F1 compared to 0.7677 for ICL Relevant (1 shot) under depth-$k$ pooling, representing a 49\% relative improvement. This confirms that even minimal guidance about topic-specific relevance criteria can substantially improve LLM judgement quality, particularly for long and complex documents.

When compared against human-only pooling baselines (top-3 depth-$k$ judgements), LLM-based approaches are competitive in MCC on DL-19 and DL-20, and they remain below the strongest human baseline on TREC-8. This pattern highlights the persistent challenge of long-document assessment while also indicating that hybrid pooling with LLMs is already practical for many realistic evaluation scenarios.

\begin{figure}
    \centering
    \includegraphics[width=\linewidth]{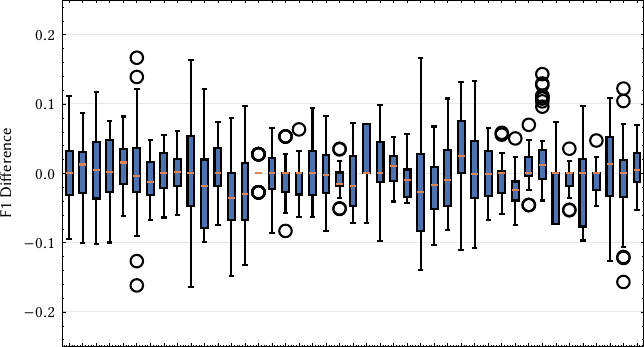}
    \Description{Per-query differences in F1 scores across ten executions of the ICL 1-shot setting on TREC-DL 2019. Each boxplot represents the distribution of pairwise F1 differences for a given query, obtained by randomly sampling one document as an in-context example in each execution.}
    \caption{Per-query differences in F1 scores across ten executions of the ICL 1-shot setting on TREC-DL 2019. Each boxplot represents the distribution of pairwise F1 differences for a given query, obtained by randomly sampling one document as an in-context example in each execution.}
    \label{fig:boxplots}
\end{figure}

 Finally, Figure~\ref{fig:boxplots} illustrates the variability of F1 scores at the query level under the ICL 1-shot configuration. For each query, ten independent executions were performed, where one in-context example was randomly selected from the pool of documents. The box-and-whisker plots show the distribution of pairwise differences in F1 scores across these executions. The results reveal substantial variability for many queries, indicating that performance is highly sensitive to the particular choice of in-context examples. This level of instability is concerning from a reliability perspective, as it suggests that dataset construction or evaluation outcomes may vary significantly depending on arbitrary sampling decisions rather than systematic methodological factors.

\subsection{Discussion}

Our findings address two central challenges in LLM-based relevance assessment: how to allocate scarce human judgements and how to condition LLMs effectively under strict context and cost constraints.

\textbf{How should human judgements be allocated in hybrid human--LLM settings?}
Our results show that concentrating human effort on shallow, top-ranked documents through depth-$k$ pooling is consistently more effective than stratified sampling. This extends long-standing insights from IR evaluation to hybrid human--LLM workflows. In contrast, stratified sampling dilutes supervision quality by including large numbers of low-value non-relevant documents and relies on assumptions about relevance class distributions that are unavailable in realistic evaluation scenarios. These findings indicate that effective hybrid pooling strategies should prioritise judgement quality over uniform coverage when annotation budgets are limited.

\textbf{What are the limitations of standard in-context learning for relevance assessment?}
With respect to LLM prompting, our analysis reveals fundamental limitations of standard ICL. While selecting relevant examples is preferable to random selection, increasing the number of examples does not reliably improve performance and can be detrimental for collections with long documents due to attention and context window constraints. This highlights a key limitation of example-heavy prompting: instance-level supervision does not scale gracefully with document length or annotation volume.

\textbf{Given these limitations, can relevance be modelled more explicitly and efficiently?}
The answers to the previous questions motivate a shift away from example-centric prompting towards approaches that abstract human supervision into more compact representations of relevance. In particular, they raise the question of whether relevance criteria can be explicitly modelled in a form that reduces prompt length, improves robustness to document length, and remains interpretable. In the following section, we investigate this question by introducing and evaluating Relevance Context Learning, which operationalises relevance as a concise, topic-specific narrative.


\section{Relevance Context Learning}
\label{sec:rcl}

Results from the previous section revealed several important limitations of standard in-context learning for relevance assessment. Beyond the fact that performance does not consistently improve as more examples are added, and may even degrade beyond a certain point, we also observed substantial variability at the per-query level depending on the specific examples selected. This sensitivity to arbitrary in-context example choices raises concerns about the reliability and reproducibility of the method. While such behaviour might be partially tolerable for collections like DL-19 and DL-20, where documents are short passages, it becomes particularly problematic for collections such as TREC-8, where documents are substantially longer. In these scenarios, providing multiple full-length document examples quickly exhausts the LLM’s attention span, further limiting the number of examples that can be effectively used and exacerbating instability in the results.

This observation motivates a fundamental question: rather than providing the LLM with multiple isolated query-document examples, can we distill the essential relevance criteria from human judgements into a more compact and generalizable form? To address this, we propose Relevance Context Learning (RCL), a method that generates explicit relevance narratives and assessment instructions from human-judged documents using a new structured form of instruction induction~\cite{Honovich2023} for relevance. These narratives aim to condense the most important aspects of what constitutes relevance for a given topic, while the instructions provide guidance for a downstream LLM to judge new documents without relying on traditional in-context examples.

\subsection{Method}
\label{sec:rcl-method}

RCL operates in two stages. First, we use an \textit{instructor} LLM to analyse the human-judged documents for a topic and generate an explicit relevance narrative that describes what makes a document relevant or non-relevant for that particular information need. Second, we use this generated narrative as a prompt to guide a second, assessor LLM (which may be the same model) when assessing new query-document pairs.

\subsubsection{Stage 1: Relevance Narrative Generation}

Given a set of human-judged query-document pairs $R_{human} = \{(q, d_i, r_i)\}$ for a topic $q$, we prompt an LLM to generate a relevance narrative $N_q$. This narrative is a natural language description that captures the underlying relevance criteria for the topic. We explore three variants for generating narratives:

\begin{itemize}
    \item \textbf{RCL with all human-judged}: The LLM receives all human-judged documents (both relevant and non-relevant) and is asked to describe what distinguishes relevant from non-relevant documents for the topic.
    \item \textbf{RCL with relevant only}: The LLM receives only the relevant human-judged documents and is asked to describe what makes these documents relevant.
    \item \textbf{RCL with non-relevant only}: The LLM receives only the non-relevant human-judged documents and is asked to describe what makes these documents non-relevant.
\end{itemize}

The generated narrative $N_q$ serves as a compact representation of the topic's relevance criteria, distilling information from potentially many examples into a single, coherent description.

\begin{figure}
    \centering
    \includegraphics[width=\linewidth]{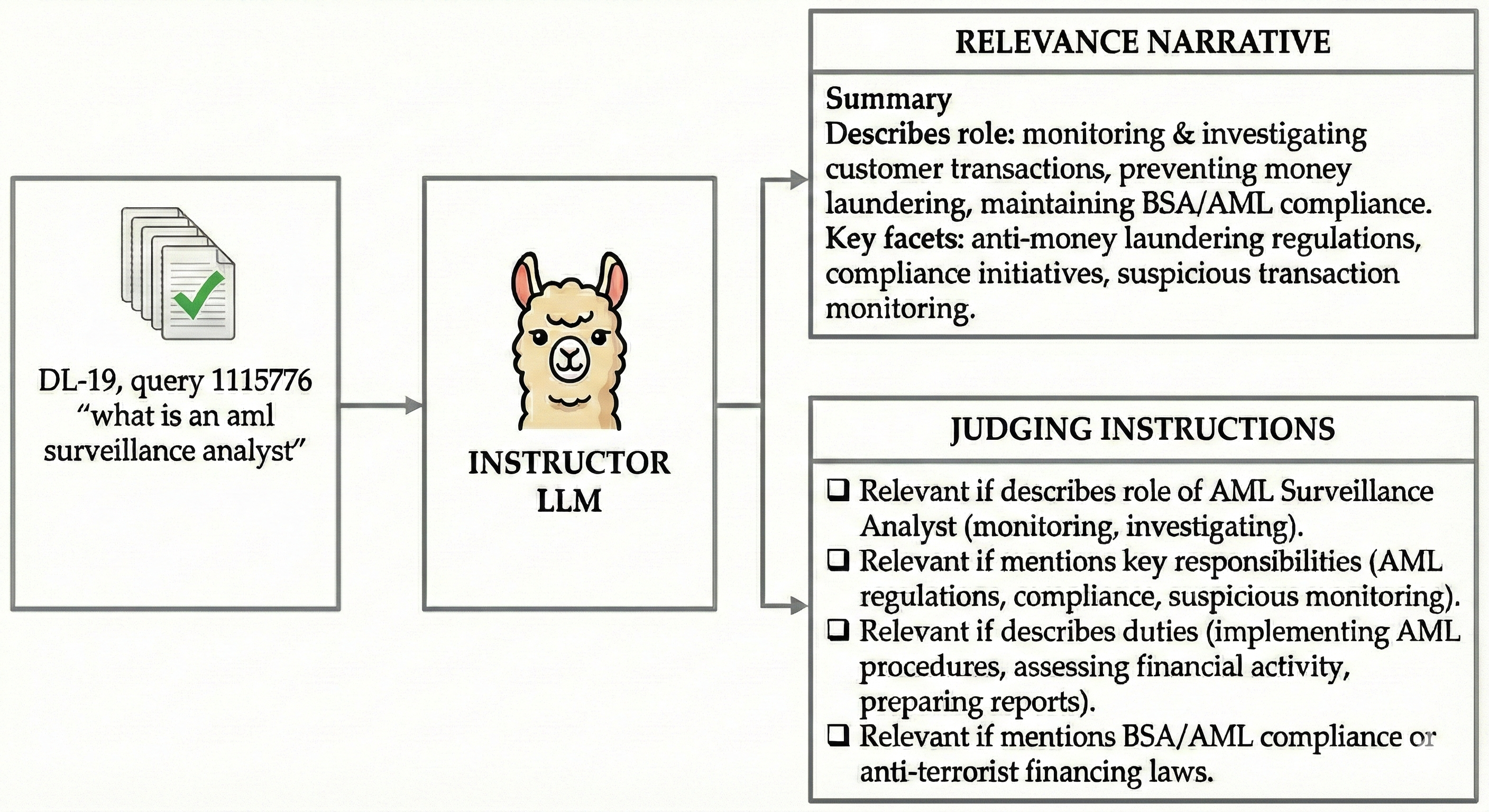}
    \Description{Schematic overview of narrative generation: the Instructor LLM processes the input query and a set of judged documents to synthesise a high-level Relevance Narrative and granular Judging Instructions. The example displays the output for TREC DL-19 query 1115776 ("what is an aml surveillance analyst") when including only relevant documents.}
    \caption{Schematic overview of narrative generation: the Instructor LLM processes the input query and a set of judged documents to synthesise a high-level Relevance Narrative and granular Judging Instructions. The example displays the output for TREC DL-19 query 1115776 ("what is an aml surveillance analyst") when including only relevant documents.}
    \label{fig:rcl-example}
\end{figure}

In \Cref{fig:rcl-example}, we include an example of how this first stage works for a particular query of the DL-19 collection. We provide the LLM with a set of documents and their corresponding human judgements, in this example, only relevant judgements. The LLM generates a narrative that condenses what makes a document relevant for this query and a set of detailed instructions that will help the downstream assessor LLM to better judge the relevance of new query-document pairs.

\subsubsection{Stage 2: Narrative-Guided Assessment}

Once we have generated the relevance narrative $N_q$, we use it to guide the LLM when judging new query-document pairs. For each document $d$ to be assessed, we construct a prompt that includes:

\begin{enumerate}
    \item Task instructions describing the relevance assessment task
    \item The relevance narrative $N_q$ for the topic and the detailed judging instructions
    \item The query $q$ and document $d$ to be judged
\end{enumerate}

The LLM then generates a relevance judgement based on this narrative-guided prompt, without requiring additional in-context examples. This approach allows the LLM to leverage the distilled relevance criteria while avoiding the attention limitations associated with long in-context examples.

\subsection{Experimental Setup}
\label{sec:rcl-setup}

We use the same datasets (DL-19, DL-20, and TREC-8), model configuration (\texttt{Llama-3.1-8B-Instruct}), and evaluation metrics (F1 and MCC) as described in Section~\ref{sec:hybrid-pooling}. All experiments use the depth-$k$ pooling strategy with human assessors judging the top-3 documents from each participating system.

For narrative generation (Stage 1), we use the same LLM model with a carefully designed prompt that instructs the model to analyse the human-judged documents and produce a concise description of relevance criteria. The prompt asks the LLM to identify patterns across the judged documents and articulate what distinguishes relevant from non-relevant content. The model is instructed to focus on key themes, required information types, and relevance boundaries rather than simply summarising individual documents. 

The generated narratives are then stored and reused for all documents that need to be judged for that topic. This one-time generation cost is amortized across all subsequent judgements for the topic, making narrative generation highly efficient compared to repeatedly providing in-context examples. For a typical topic with hundreds of documents to judge, the narrative generation overhead represents less than 1\% of the total inference cost.

For assessment (Stage 2), we construct prompts that begin with task instructions, followed by the relevance narrative, and ending with the query-document pair to be judged. The narrative serves as explicit guidance that helps the LLM understand what the information need entails and what types of content would satisfy it. This approach differs fundamentally from ICL, where the LLM must implicitly infer relevance patterns from examples.

\subsection{Results}
\label{sec:rcl-results}

\begin{table*}
\centering

\caption{Comparison of RCL and ICL strategies on DL-19, DL-20 and TREC-8. Best values for each collection and metric are shown in \textsc{bold}, while second best are \un{underlined}.}
\label{tab:rcl}

\begin{tabular}{l|c|cc|cc|cc}

\toprule

\multicolumn{2}{l}{} & \multicolumn{2}{|c|}{\textbf{DL-19}} & \multicolumn{2}{c|}{\textbf{DL-20}} & \multicolumn{2}{c}{\textbf{TREC-8}} \\

\midrule

\multicolumn{1}{l}{\textbf{Strategy}} & \textbf{Shots} & \textbf{F1} & \textbf{MCC} & \textbf{F1} & \textbf{MCC} & \textbf{F1} & \textbf{MCC} \\

\midrule

\textsc{Zero Shot} & 0 & 0.7771 & 0.6092 & 0.6869 & 0.6754 & 0.5204 & 0.5096  \\

\midrule

\textsc{TREC Narrative} & 0 & - & - & - & - & 0.7658 & 0.7212  \\

\midrule

\textsc{RCL} w/ all human-judged & 0 & 0.8680 & 0.8834 & 0.8205 & \textbf{0.8821} & \un{0.8013} & \textbf{0.7921} \\
\textsc{RCL} w/ all human-judged relevants & 0 & \un{0.8770} & \textbf{0.9087} & 0.8237 & \un{0.8786} & \textbf{0.8267} & 0.7850 \\
\textsc{RCL} w/ all human-judged non-relevants & 0 & 0.8570 & \un{0.8841} & 0.8046 & 0.8530 & 0.7842 & 0.7842 \\

\midrule

\textsc{ICL} w/ all human-judged & - & \textbf{0.8890} & 0.8702 & \textbf{0.8767} & 0.8781 & 0.7468 & 0.7181 \\
\textsc{ICL} w/ all human-judged relevants & - & 0.8736 & 0.8702 & \un{0.8467} & 0.8753 & 0.6541 & 0.6394 \\
\textsc{ICL} w/ all human-judged non-relevants & - & 0.8143 & 0.8283 & 0.7818 & 0.7479 & 0.7848 & \un{0.7857} \\

\bottomrule

\end{tabular}
\end{table*}

\begin{table*}
\centering

\caption{Results of RICL (RCL + ICL) experiment on the DL-19, DL-20 and TREC-8 datasets. In this experiment, we used the RCL strategy that uses only relevant documents. Best values for each collection and metric are shown in \textsc{bold}, while second best are \un{underlined}.}
\label{tab:rcl-vs-icl}

\begin{tabular}{l|c|cc|cc|cc}

\toprule

\multicolumn{2}{l}{} & \multicolumn{2}{|c|}{\textbf{DL-19}} & \multicolumn{2}{c|}{\textbf{DL-20}} & \multicolumn{2}{c}{\textbf{TREC-8}} \\

\midrule

\multicolumn{1}{l}{\textbf{Strategy}} & \textbf{Shots} & \textbf{F1} & \textbf{MCC} & \textbf{F1} & \textbf{MCC} & \textbf{F1} & \textbf{MCC} \\

\midrule

\textsc{RCL} w/ all human-judged relevants & 0 & \textbf{0.8770} & \textbf{0.9087} & 0.8237 & 0.8786 & \textbf{0.8267} & \un{0.7850} \\

\midrule

\multirow{3}{*}{\textsc{RICL}}
& 1 & 0.8331 & 0.8743 & 0.8249 & \un{0.9000} & 0.7558 & 0.7074 \\
& 2 & \un{0.8383} & 0.8881 & 0.8416 & 0.8931 & 0.7641 & 0.7032 \\
& 3 & 0.8367 & 0.8819 & 0.8437 & 0.8875 & 0.7708 & 0.7173 \\

\arrayrulecolor{black!30}
\midrule
\arrayrulecolor{black}

\multirow{3}{*}{\textsc{RICL Relevant}}
& 1 & 0.8311 & 0.8852 & 0.8452 & 0.8928 & \un{0.8174} & \textbf{0.7886} \\
& 2 & 0.8234 & 0.8828 & \un{0.8468} & \textbf{0.9073} & 0.8067 & 0.7767 \\
& 3 & 0.8203 & \un{0.8907} & \textbf{0.8487} & 0.8986 & 0.7933 & 0.7646 \\

\bottomrule

\end{tabular}
\end{table*}

We present results from two experiments that explore the capabilities of RCL: first comparing RCL against standard ICL, and then investigating the combination of narratives with in-context examples, RICL (RCL + ICL).

\subsubsection{RCL vs ICL}

\Cref{tab:rcl} compares RCL and ICL using all available human-judged documents. The key distinction is that RCL operates in a zero-shot setting, relying solely on a generated relevance narrative, whereas ICL conditions the LLM on all human-judged examples directly. This comparison isolates the effect of explicit relevance narratives relative to example-based prompting.

\textbf{Overall performance.}
Across DL-19 and DL-20, RCL is competitive with standard ICL and yields consistently strong MCC values. On DL-19, the highest F1 is obtained by ICL with all human-judged documents (0.8890), while RCL with relevant documents achieves the highest MCC (0.9087). On DL-20, ICL with all human-judged documents again yields the highest F1 (0.8767), whereas RCL with all human-judged documents attains the highest MCC (0.8821). Overall, the results indicate a trade-off in which ICL is often stronger on raw F1 for short-document collections, while RCL remains highly competitive and can offer better discriminative reliability.

\textbf{Impact of document length.}
The largest performance differences arise on TREC-8, where document length becomes critical. RCL with relevant documents reaches 0.8267 F1, substantially above ICL with relevant documents (0.6541) and ICL with all human-judged documents (0.7468). Relative to ICL with relevant documents, this corresponds to a 26\% F1 improvement. This confirms that compact relevance narratives are substantially more effective than full-document examples when context budgets are constrained. Notably, the human-written TREC narrative also performs strongly (0.7658 F1), though it remains inferior to RCL narratives derived from judged documents.

\textbf{Narrative construction choices.}
We observe that the documents used to generate relevance narratives influence performance. On DL-19, narratives derived solely from relevant documents yield the best results, while narratives generated from non-relevant documents remain competitive. This suggests that both positive and negative evidence can inform relevance criteria, although relevant-only narratives provide the clearest signal. On DL-20, differences across narrative construction strategies are smaller, indicating robustness to this design choice.

\textbf{Zero-Shot Baseline.}
Zero-shot prompting performs substantially worse than both RCL and ICL across all collections. On DL-19 and TREC-8, the best RCL setting improves F1 by about 13\% and 59\%, respectively, demonstrating that topic-specific guidance is essential, particularly for long and complex documents.

Beyond effectiveness, the hybrid approach also offers increased robustness to evaluation bias with respect to zero-shot and sampling. Balog et al.~\cite{Balog2025} show that LLM-based judges can exhibit bias towards neural rankers when assessment relies on model-generated signals. In our setting, relevance narratives are induced from a shallow depth-$k$ pool that aggregates outputs from diverse retrieval systems, encouraging the Instructor LLM to learn ranker-agnostic relevance criteria. This diversity reduces the risk of encoding ranker-specific biases when LLMs are used to assess deeper ranks.

\subsubsection{Combining Narratives and Examples (RICL)}

We next examine whether combining relevance narratives with in-context examples further improves performance. We refer to this hybrid approach as Relevance and In-Context Learning (RICL). In RICL, prompts consist of task instructions, the relevance narrative, a small number of in-context examples, and the document to be judged.

\Cref{tab:rcl-vs-icl} reports results for RICL using different example selection strategies and numbers of shots.

\textbf{Does combining help?}
The effectiveness of RICL depends on the collection. On DL-19, RCL alone already captures sufficient relevance information, and adding examples does not improve performance. In contrast, on DL-20, RICL yields consistent gains over RCL, with improvements of up to 3.0\% in F1 (0.8487 vs.\ 0.8237) and 3.3\% in MCC (0.9073 vs.\ 0.8786). This indicates that for some collections, narratives and examples provide complementary information.

\textbf{Example selection.}
When relevance narratives are provided, example selection remains important but exhibits different patterns than pure ICL. On DL-20, RICL with randomly selected examples achieves the highest F1, while selecting only relevant examples yields the best MCC, suggesting that narratives help define relevance boundaries and allow both positive and negative examples to contribute meaningfully.

\textbf{Number of Examples.}
RICL benefits from a small number of examples. Performance generally improves from one to two or three shots, after which gains plateau. Although narratives mitigate some of the attention limitations observed in standard ICL, diminishing returns remain beyond 2--3 examples.

\subsection{Discussion}
\label{sec:discussion}

RCL investigates whether explicit modelling of topic-level relevance criteria can improve the use of Large Language Models (LLMs) for relevance assessment under realistic evaluation constraints. Across multiple collections and experimental settings, our findings highlight the advantages of abstracting human supervision into compact relevance narratives and clarify when example-based prompting remains beneficial.

\textbf{Can Relevance Context Learning outperform standard ICL?}
Our results show that Relevance Context Learning (RCL) is strongly competitive with standard In-Context Learning (ICL), despite operating in a zero-shot setting during assessment. This finding is particularly significant given that ICL has been the dominant paradigm for conditioning LLMs on relevance judgements. The clearest gains appear on collections with long documents, such as TREC-8, where RCL with relevant documents improves F1 by 26\% over the corresponding ICL relevant setting (0.8267 vs.\ 0.6541). These gains arise because relevance narratives provide a concise representation of relevance criteria that avoids exhausting the LLM’s attention budget, a limitation that severely restricts the effectiveness of full-document examples in ICL. Even on collections with shorter documents, where ICL can lead in F1, RCL remains competitive and often provides stronger MCC.

\textbf{What supervision is most effective for narrative generation?}
We find that the documents used to generate relevance narratives meaningfully affect performance. Narratives synthesised from relevant-only documents generally yield the strongest results across collections. This aligns with the intuition that positive evidence provides the clearest signal for defining relevance criteria, while non-relevant documents, although informative in some cases, introduce additional noise. That said, narratives generated from non-relevant or mixed sets remain competitive, indicating that the Instructor LLM is capable of extracting useful constraints from both positive and negative evidence. In practice, however, using only relevant documents offers a strong and reliable default.

\textbf{Do narratives and examples provide complementary benefits?}
The effectiveness of combining relevance narratives with in-context examples (RICL) depends on collection characteristics. On collections with short documents, such as DL-20, RICL yields modest but consistent improvements over RCL alone, with gains of up to 2--3\% in F1 and MCC. This suggests that in such settings, examples can complement narratives by resolving residual ambiguity in relevance criteria. In contrast, on collections where narratives already capture the relevance context effectively (e.g., DL-19), or where document length severely limits context budgets (e.g., TREC-8), adding examples provides little benefit and can even be impractical. These results indicate that while narratives and examples can be complementary, narratives alone are often sufficient and more robust across settings.

\textbf{Implications for test collection construction.}
Taken together, these findings support a shift from example-heavy prompting towards explicit relevance modelling for LLM-based relevance assessment. RCL offers several practical advantages: it reduces prompt length and inference cost, scales naturally to long documents, and produces interpretable relevance descriptions that can be inspected or refined by humans. When annotation budgets are limited, and especially when document length is large, RCL provides a principled and efficient alternative to standard ICL. For practitioners working with short documents and ample computational resources, RICL can offer incremental improvements, but its benefits are context-dependent.

From a practical perspective, our framework also allows quality improvements to be achieved selectively and cost-effectively. In particular, since relevance narratives are generated once per query, upgrading the \emph{Instructor} LLM can improve assessment quality with minimal additional cost. In preliminary experiments on DL-19, we observed that replacing the Instructor LLM with a larger and more capable model (\texttt{Llama-4-Scout-16B-16E-Instruct}), while keeping the same Assessor LLM (\texttt{Llama-3.1-8B-Instruct}), improved F1 for the RCL Relevant setting from 0.8680 to 0.8809. This suggests that investing computational resources in narrative generation, rather than per-document assessment, offers a favorable cost--quality trade-off.

Overall, this work demonstrates that transforming human judgements into explicit, reusable relevance narratives is a more scalable and effective way of conditioning LLMs for relevance assessment than relying solely on raw in-context examples. We believe this paradigm has broad implications for the construction of future IR evaluation datasets and for the design of human--LLM hybrid assessment workflows.


\section{Related Work}
\label{sec:rw}

The Cranfield paradigm~\cite{Sparck1975,Cleverdon1967} established the foundation for systematic IR evaluation through reusable test collections comprising documents, queries, and relevance judgements. Given the impracticality of exhaustively judging all documents for each query, pooling strategies emerged as the de facto standard for building test collections~\cite{Voorhees2005}. Traditional depth-$k$ pooling collects the top-$k$ documents from multiple participating systems and submits them for human judgement~\cite{Voorhees2001a}. While this approach has proven effective for decades, it introduces potential biases towards systems that contributed to the pool~\cite{Buckley2007}.

Numerous variations on pooling have been proposed to address these limitations. Stratified sampling selects documents proportionally across ranks~\cite{Lipani2019}, while move-to-front pooling prioritises unjudged documents that appear at high ranks~\cite{Cormack1998}. Statistical methods have been developed to evaluate systems using incomplete judgements~\cite{Aslam2006}, and research has examined the reliability and bias properties of various pooling strategies~\cite{Zobel1998,Sanderson2005,Lipani2016,Losada2016a}. Despite these advances, pooling-based approaches remain labour-intensive and costly when building large-scale test collections.

The emergence of Large Language Models has introduced new possibilities for automating relevance assessment. Recent studies have investigated whether LLMs can serve as substitutes or complements to human assessors. Several works have demonstrated that LLMs can achieve moderate to high correlation with human judgements~\cite{Upadhyay2024c,Thomas2024}. These studies typically employ zero-shot prompting, where the model is simply instructed to assess relevance, or few-shot ICL, where a small number of labelled examples are provided in the prompt.

However, LLM-based assessors exhibit several limitations. They can display biases towards certain document characteristics~\cite{Soboroff2024}, show inconsistency across different query types~\cite{Otero2025b}, and may not generalise well across collections. Alaofi et al.~\cite{Alaofi2026} examine the gullibility of LLMs, demonstrating that they can be easily misled by the presence of query terms even in irrelevant passages. Their analysis reveals that, although LLMs achieve high correlation with human judgements, they exhibit lower discriminative power and may introduce system bias by favouring traditional retrieval methods over neural approaches. Recent work has also explored ensemble approaches that combine judgements from multiple LLMs~\cite{Faggioli2023} and has studied the reliability of LLM-based assessments for constructing reusable test collections~\cite{Abbasiantaeb2024}.

In parallel, in-context learning (ICL) has emerged as an approach for improving LLM-based relevance assessment. ICL provides the LLM with a small number of query–document pairs and their relevance labels as demonstrations before requesting a judgement on new pairs~\cite{McKechnie2025a}.

A fundamental limitation of standard ICL is that it treats examples as isolated instances without explicitly capturing the underlying task criteria. As a result, the model must implicitly infer the task objective from individual examples, which, in the context of relevance assessment, can lead to inconsistent or superficial representations of topic-specific relevance. In other fields, recent work has explored instruction induction as a means of making latent task definitions explicit. Rather than relying solely on demonstrations, these approaches prompt LLMs to infer an explicit natural-language instruction that explains the examples, effectively externalizing the task description~\cite{Honovich2023,Kharrat2025,Xiao2025}. These studies show that LLMs can successfully induce task-level instructions from a small set of examples, achieving performance comparable to standard ICL while improving efficiency and consistency.


\section{Conclusions}
\label{sec:conclusions}

In this work, we addressed two central challenges in the use of large language models for relevance assessment in the construction of IR test collections. First, we proposed a practical hybrid pooling strategy that allocates annotation effort between human assessors and LLMs. The depth-\textit{k} pooling scheme, in which human assessors judge only the highest-ranked documents while LLMs are applied to deeper ranks, achieves competitive evaluation reliability while substantially reducing human annotation cost. Experiments on three standard collections show that this strategy offers a realistic and effective alternative to fully human-judged pools.

Second, we introduced Relevance Context Learning (RCL), a narrative-based approach that derives explicit relevance criteria from human judgements rather than relying on traditional in-context examples. RCL addresses a key limitation of standard in-context learning, namely performance saturation or degradation as more examples are added—an issue that becomes particularly severe for collections with long documents due to context window constraints. By distilling relevance judgements into compact, topic-specific narratives, RCL achieves performance comparable to or exceeding that of ICL while requiring no in-context examples during assessment. This advantage is especially pronounced on collections such as TREC-8, where narrative-based prompting markedly outperforms example-based approaches, highlighting the robustness of RCL under strict context and cost constraints.

Beyond efficiency and effectiveness, the proposed hybrid framework also provides a promising direction for mitigating emerging risks associated with large-scale LLM-based evaluation, including model bias and model collapse. By anchoring relevance narratives in human judgements drawn from diverse retrieval systems, RCL reduces reliance on self-referential LLM signals and helps prevent feedback loops in which LLM-based rankers and judges reinforce shared biases. Explicit relevance narratives further act as a stabilizing abstraction, preserving human-defined relevance criteria even as underlying models evolve.

While these results are encouraging, they also point to several directions for future work. First, the quality of relevance narratives is inherently tied to the capabilities of the underlying Instructor LLM and the amount of human supervision available. Exploring the use of larger or more specialized models (e.g., fine-tuned or domain-adapted models) for narrative generation may further improve consistency, coverage, and faithfulness to human judgements. Second, although our study focuses on binary relevance labels, the narrative-based formulation of RCL naturally extends to more expressive relevance schemes, such as graded relevance or aspect-based relevance criteria. Investigating how narratives can capture multiple degrees or dimensions of relevance represents a promising avenue for extending RCL beyond binary assessment.

Two additional limitations warrant dedicated analysis. First, our current experiments do not isolate how the specific set of human-judged relevant documents affects narrative quality. A useful next step is to generate multiple narratives from different subsets of relevant documents for the same topic and measure their similarity, stability, and downstream assessment effectiveness. This would clarify whether RCL depends on particular relevant examples or captures a robust topic-level notion of relevance. Second, the trade-off between human and LLM judgement effort remains only partially characterised. In particular, it is important to quantify how performance changes as the number of human-judged relevant documents varies, and to identify operating points that best balance annotation cost against evaluation reliability.

In addition, understanding the robustness of RCL to noisy or sparse human labels remains an important open question. Techniques for uncertainty estimation, narrative validation, or iterative refinement could help mitigate potential biases introduced during narrative generation. Finally, extending RCL to support dynamic or evolving relevance criteria, such as those arising in continuously updated collections, represents a promising direction for future research.

Overall, the methods presented in this paper provide a strong foundation for more efficient, scalable, and interpretable test collection construction using LLMs, while maintaining the level of reliability required for rigorous IR system evaluation and offering a principled path towards more robust human--LLM hybrid assessment workflows.

\section*{Declaration on the use of Generative AI}

During the preparation of this work, the authors used generative AI to assist in the writing of this manuscript and in the writing of the code used to run the experiments. The authors carefully reviewed and edited the content produced by the generative AI tools, and take full responsibility for the content of this publication.

\begin{acks}
All authors acknowledge funding from the Ministry of Science, Innovation and Universities of the Government of Spain (project PID2022-137061OB-C21, MCIN/AEI/10.13039/501100011033), as well as from the Department of Education, Science, Universities, and Vocational Training of the Xunta de Galicia (grant GRC ED431C 2025/49). CITIC, as a center accredited for excellence within the Galician University System and a member of the CIGUS Network, receives subsidies from the Department of Education, Science, Universities, and Vocational Training of the Xunta de Galicia. Additionally, it is co-financed by the EU through the FEDER Galicia 2021-27 operational program (Ref. ED431G 2023/01).
\end{acks}

\printbibliography

\end{document}